\documentclass[12pt]{iopart}

\usepackage{bm,graphicx}

\begin{document}

\title[CF of bosons in rapidly rotating atomic traps]{Composite fermion theory of rapidly rotating two-dimensional bosons}

\author{N. Regnault$^1$, C.C. Chang$^2$,  Th. Jolicoeur$^{1,3}$, J.K. Jain$^2$}

\address{$^1$ Laboratoire Pierre Aigrain, D\'epartement de Physique, 24, rue Lhomond, 75005 Paris, France.}

\address{$^2$ Department of Physics, 104 Davey Laboratory, The Pennsylvania State University, Pennsylvania 16802, U.S.A.}

\address{$^3$ Laboratoire de Physique Th\'eorique et Mod\`eles Statistiques, Universit\'e Paris-Sud, 91405 Orsay, France.}

\begin{abstract}
Ultracold neutral bosons in a rapidly rotating atomic trap have been predicted to exhibit fractional quantum Hall-like states. We describe how the composite fermion theory, used in the description of the fractional quantum Hall effect for electrons, can be applied to interacting bosons. Numerical evidence supporting the formation of composite fermions, each being the bound state of a boson and one flux quantum, is shown for filling fractions of the type $\nu=p/(p+1)$, both by spectral analysis and by direct comparison with trial wave functions.  The rapidly rotating system of two-dimensional bosons thus constitutes an interesting example of ``statistical transmutation," with bosons behaving like composite  fermions.  We also describe the difference between the electronic and the bosonic cases when $p$ approaches infinity. Residual interactions between composite fermions are attractive in this limit, resulting in a paired composite-fermion state described by the Moore-Read wave function.

\end{abstract}

\maketitle

\section{Introduction}

Under appropriate circumstances, strong short range repulsion between bosons can mimic the Pauli exclusion principle of fermions. This has been known for a long time in the case of bosons in one dimension \cite{Girardeau60}, and the Lieb-Liniger criterion \cite{Lieb63} discriminates in which case this fermionization occurs. Experiments based on Bose-Einstein condensates of atomic gases allow a test of the concept of statistical transmutation of interacting bosons into fermions.  Such behavior has been recently observed in experiments \cite{Kinoshita04,Paredes04}.

Here we discuss mapping of bosons into a different kind of fermions, called composite fermions.  Each composite fermion is the bound state of a boson and one quantized vortex of the wave function, often conveniently though of as the bound state of a boson and one flux quantum, $\phi_0=hc/e$.  Composite fermions are expected to occur when ultracold bosons are placed in a rapidly rotating atomic trap. Such rotation  leads to a variety of rich phenomena, such as the Abrikosov lattice of vortices \cite{Madison00,Aboshaeer01}. For high rotation frequency, it has been predicted \cite{Wilkin98,Wilkin00,Cooper01,Regnault03} that bosons, under appropriate conditions, will behave like two dimensional electron systems (2DES) in the fractional quantum Hall effect (FQHE) regime. [More precisely:  A strong confinement along the rotation axis is required so the system can be considered as two dimensional. A fine tuning between the rotation frequency and the harmonic trap frequency must be reached so that the centrifugal force compensate the harmonic trapping force.  Further, the interaction between bosons has to be weak enough so that any mixing with higher-lying Landau levels is negligible.  Thus the system of neutral bosons subjected to a Coriolis force is equivalent to a system of charged particles in an external magnetic field.  The filling factor $\nu$ is the ratio of the number $N$ of particles over the number $N_V$ of quantized vortices that would be present in the system if it were a Bose condensate.]  This system of bosons in the lowest Landau level (LLL) is an ideal playground where theories of the FQHE can be tested, and extended into new regimes.  In some respects, the bosonic system offers more flexibility than the electronic one.  The strength of the interaction can be tuned using the Feschbach resonance, and there are many choices for the condensed atoms; that allows the possibility of studying systems with different types of interactions (delta function ~\cite{Wilkin98}, hollow core~\cite{Regnault04} or dipole-dipole~\cite{Baranov05}) and quantum statistics (Bose or Fermi).

For the {\em electron} system in the lowest Landau level, Laughlin's wave function\cite{Laughlin83} provides a good approximation for the ground state at filling factor $1/3$. However,  a larger class of fractions of the type $p/(2p\pm 1)$ appear in experiments, and  are understood through the formation of composite fermions (CFs)~\cite{Jain89}, namely electrons bound to an {\em even} number of quantized vortices.  The problem of interacting electrons maps onto free composite fermions in an effective magnetic field, and the fractions $p/(2p\pm 1)$ are explained as the integral quantum Hall effect of composite fermions. 

In this paper we focus on the case of bosonic atoms with only one hyperfine species (spinless bosons). In this case, the interaction between atoms occurs through s-wave scattering.  Such a system has the property that the Laughlin wave function\cite{Laughlin83} is the \textit{exact} ground state at filling factor $\nu=1/2$~\cite{Wilkin98}. 
However, one can ask:  Do composite fermions also occur in rotating two-dimensional bose systems?  The composite fermionization of the bosonic system will have many testable consequences beyond Laughlin's wave function, most important being the prediction of incompressible states at many fractions of the type $\nu=p/(p\pm 1)$. Such fractions can also be deduced from the fermion-boson mapping based on a gauge field approach\cite{Wilczek90,Xie91,Quinn01}. We show here numerical evidence of the accuracy of the CF description by identifying the incompressible states,  by spectral analysis, and by direct comparison of the Jain's CF wave functions with exact wave functions.   We shall use the spherical geometry to establish detailed and quantitative correspondence between interacting bosons and free fermions. This study extends previous works on disk\cite{Viefers00,Wilkin00,Cooper99} and toroidal geometries\cite{Cooper01}.  An advantage of  the spherical geometry over the disk geometry is in facilitating the identification of incompressible states of the type 2/3 and 3/4.

In section \ref{cheffective}, we present the effective Hamiltonian that describes the system. Section \ref{chlaughlin} is devoted to the Laughlin wave function, which is the exact ground state of the rotating bosons interacting through s-wave scattering at filling factor $\nu=1/2$. We present the spherical geometry in section \ref{chsphere}, which provides a convenient framework for numerical calculations. In section \ref{chcf}, we give an overview of the CF theory and its extension to bosons. Section \ref{chnumerical} shows numerical evidence for the presence of incompressible states at several fillings, and demonstrates the suitability of the CF theory for their description.  Finally in section \ref{chpfaffian}, we discuss the state at $\nu=1$, which requires a consideration of pairing of composite fermions, described by the Moore-Read's Pfaffian wave function.

\section{Effective Hamiltonian}\label{cheffective}

In the rotating frame, the Hamiltonian that describes $N$ atoms of mass $m$ in a rotating harmonic trap is given by~:
\begin{eqnarray}
H&=&\sum_{i=1}^N \frac{1}{2m}\left({\bf p}_i - m{\bf\hat{z}}\times{\bf r}_i\right)^2 + \frac{m}{2}\left[\left(\omega_r^2-\omega^2\right)(x_i^2+y_i^2)+\omega_z^2 z_i^2\right]\nonumber\\
&&+\sum_{i<j} V\left({\bf r}_i-{\bf r}_j\right),
\label{hamiltonian3D}
\end{eqnarray}
where $\omega_r$ is the radial harmonic trap frequency, $\omega_z$ is the axial harmonic trap frequency and $\omega$ the trap rotation frequency (rotation along the ${\bf\hat{z}}$ axis). For ultracold bosonic gases, the interaction is dominated by $s$-wave scattering. Thus the interaction can be approximated by a delta function~:
\begin{eqnarray}
V\left({\bf r}\right)&=&\frac{4\pi \hbar^2 a_s}{m}\delta^{(3)}\left({\bf r}\right),
\label{interaction3D}
\end{eqnarray}
with $a_s$ the $s$-wave scattering length.

We will now assume the critical rotation frequency $\omega=\omega_r$. Thus in the plane perpendicular to the rotation axis, the system is equivalent to charged particles in a magnetic field along the ${\bf\hat{z}}$ axis with cyclotron frequency $\omega_c=2\omega$. We also require strong confinement along the ${\bf\hat{z}}$ axis, so the wave function in this direction is the ground state of the harmonic oscillator along ${\bf\hat{z}}$. The effective 2D interaction is given by~:
\begin{eqnarray}
V\left({\bf r}\right)=g l^2\delta^{(2)}\left({\bf r}\right)&\;\;{\rm with}\;\;&g=\sqrt{32\pi}\;\hbar \omega \frac{a_s}{l_z},
\label{interaction2D}
\end{eqnarray}
where ${\bf r}$ is now the particle coordinate in the $x-y$ plane, $l_z=\sqrt{\hbar/m\omega_z}$ is the characteristic length of the ${\bf\hat{z}}$ axis oscillator and $l=\sqrt{\hbar/2m\omega}$ is the magnetic length. The effective Hamiltonian can be then written as~:
\begin{eqnarray}
H_{2d}&=&\sum_{i=1}^N \frac{1}{2m}\left(p_i - m{\bf\hat{z}}\times {\bf r}_i\right)^2\;+\;g l^2\sum_{i<j}\delta^{(2)}\left({\bf r}_i-{\bf r}_j\right).
\label{hamiltonian2D}
\end{eqnarray}
This two dimensional regime is formally equivalent to the FQHE for 2DES. The filling factor $\nu$ is defined as $\nu=h\rho/2m\omega$ where $\rho$ is the areal boson density. It can be also expressed as the ratio $N/N_V$ with $N_V$ the number of quantized vortices which is the analog of the number $N_\phi$ of flux quanta in 2DES systems. We will assume in the following that the interaction strength is small enough so that Landau level mixing can be neglected, and all bosons are in the LLL. The effective Hamiltonian is then reduced to the interaction term alone~:
\begin{eqnarray}
H_{\rm LLL}&=&g l^2\sum_{i<j} \delta^{(2)}\left({\bf r}_i-{\bf r}_j\right).
\label{hamiltonian2DLLL}
\end{eqnarray}
Note that due to the bosonic statistics, we can have filling factors greater that one and still stay entirely in the LLL. Such a regime is forbidden in the fermionic case.

\section{$\nu=1/2$: the Laughlin state}\label{chlaughlin}

\begin{figure}[!htbp]
\begin{center}
\includegraphics[width=15cm, keepaspectratio]{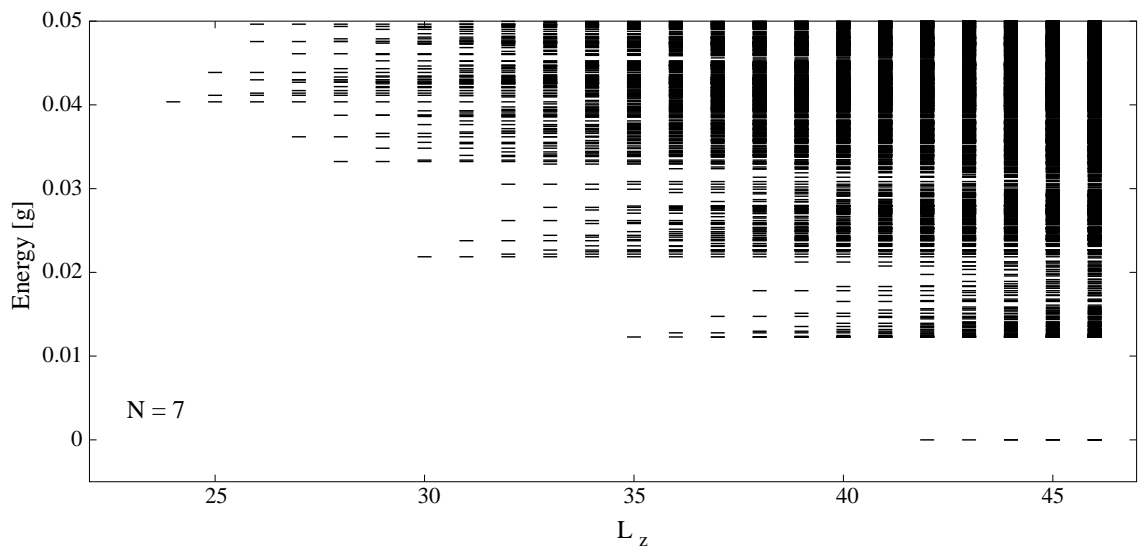}
\end{center}
\caption{Lower part of the energy spectrum for 7 bosons on the disk geometry. The Laughlin state corresponds to the zero energy ground state at $L_{z}$=42.}
\label{disk}
\end{figure}

In the LLL, the one-body eigenfunctions in the symmetric gauge are given by~:
\begin{eqnarray}
\phi_m(z)&=&\frac{1}{\sqrt{2\pi l^2 2^{\bf m} {\bf m}!}} \left(\frac{z}{l}\right)^{\bf m} e^{-\left|z\right|^2/4l^2},
\label{onebodydisk}
\end{eqnarray}
with $z=x+iy$, the positive integer ${\bf m}$ is the angular momentum of the state $(L_z=\hbar {\bf m})$ and $l$ is the magnetic length. Any N-body wave function for particles in the LLL can be expressed as a polynomial ${\cal P}$ of the $z_i$ coordinates up to a global Gaussian factor~:
\begin{eqnarray}
\Psi\left(z_1,...,z_N\right)&=&{\cal P}\left(z_1,...,z_N\right)\;\; e^{-\sum_i\left|z_i\right|^2/4l^2}.
\label{genericnbodydisk}
\end{eqnarray}
The Laughlin wave function~\cite{Laughlin83} was introduced in the case of 2DES to explain the fractional quantum Hall state at filling factor $\nu=1/3$. It describes a droplet of incompressible fluid with a mean density corresponding to this fraction. It can also be generalized for bosons~:
\begin{eqnarray}
\Psi_{{\rm Laughlin}}\left(z_1,...,z_N\right)&=&\prod_{i<j}\left(z_i - z_j\right)^2 \;\; 
e^{-\sum_i\left|z_i\right|^2/4l^2}.
\label{laughlindisk}
\end{eqnarray}
In the case of 2DES, the Laughlin wave function is a very good approximation for the true ground state. But for the system we are considering, this is the exact zero energy ground state of the effective Hamiltonian Eq.(\ref{hamiltonian2DLLL}) at filling factor $\nu=1/2$. It is also the zero energy ground state with the smallest total angular momentum (equal to  the highest filling factor). An arbitrary function built from the product of $\Psi_{{\rm Laughlin}}$ and any symmetric polynomial of the $z_i$ coordinates is also a zero energy ground state but has higher angular momentum and thus corresponds to a filling factor $\nu < 1/2$. These states represent gapless edge excitations or quasihole excitations of the 1/2 system. Quasielectrons are nucleated from the Laughlin state when reducing the number of flux quanta (or the angular momentum). Those have a nonzero gap in the thermodynamical limit \cite{Regnault03}.
These properties can be seen in exact diagonalizations. Figure \ref{disk} displays the spectrum for seven particles. The first zero energy eigenstate appears for angular momentum $L_{z}$=42 which is in agreement with Eq.~\ref{laughlindisk} $\left(L_z=N(N-1)\right)$. Higher angular momentum values also exhibits zero energy eigenstates as expected from the above discussion. 

\section{The spherical geometry}\label{chsphere}

The planar, or disk, geometry is natural for the system we are dealing with.  Unfortunately this geometry is plagued by edge effects when doing numerical calculations. Because we are interested in the bulk properties of our system, it is more suitable to work with compact geometries, such as a torus or a sphere. In this paper we will use the last one, which is easier to handle due to its $SU(2)$ symmetry, enabling a classification of the states with respect to their \textit{total} angular momentum (denoted by $L$). Let us introduce the spinor coordinates~:
\begin{eqnarray}
u_j=\cos\left(\theta_j /2\right) e^{i\phi_j/2} &\;\;{\rm and}\;\;&v_j=\sin\left(\theta_j /2\right) e^{-i\phi_j/2}. \label{spinorcoordinates}
\end{eqnarray}
The spherical geometry requires us to introduce a magnetic monopole at the center of the sphere. Its radius $R$ is then related to the number of flux quanta that pierce it~:
\begin{eqnarray}
R&=&l\sqrt{N_V/2}.
\label{radiusflux}
\end{eqnarray}
Solutions of the one-body problem are given by the monopole harmonics \cite{Wu76} (a generalization of the spherical harmonics), which take the following form in the LLL~:
\begin{eqnarray}
Y_{\bf m}(u,v)&=&\sqrt{\frac{\left(2S+1\right)!}{4\pi\left(S-{\bf m}\right)!\left(S+{\bf m}\right)!}}\;\; u^{S+{\bf m}}v^{S-{\bf m}},
\label{harmonic}
\end{eqnarray}
where ${\bf m}$ is the projection of the angular momentum along the z axis. The angular  momentum quantum number $S$ for the LLL is related to $N_V$ by the relation $2S=N_V$. In a given Landau level, the two-body interaction can be parameterized using a set of $2S+1$ number $\left\{V_{\bf m}\right\}$ called the pseudo-potentials~\cite{Haldane83}, where ${\bf m}$ is related to the \textit{relative} angular momentum between the two particles. For spinless bosons, only even-${\bf m}$ potentials are relevant. In this formalism, the delta function  interaction corresponds to all $V_{\bf m}=0$ except $V_0$. Longer range interaction can be simulated by turning on the other pseudo-potentials, the first relevant one for bosons being $V_2$.
For trial wave functions such as the Laughlin wave function, we can pass from the disk geometry to the spherical geometry using the stereographic projection~\cite{Fano86}. From a practical point of view, we drop the Gaussian factor and make the replacement~:
\begin{eqnarray}
\left(z_i-z_j\right)&\longrightarrow&\left(u_i v_j - u_j v_i\right).
\label{disktosphere}
\end{eqnarray}
Jain's wave functions are conveniently constructed directly in the spherical geometry.

\section{Composite fermions}\label{chcf}

In the case of 2DES, the CF theory~\cite{Jain89} maps the problem of strongly interacting electrons in the LLL onto weakly interacting fermions called composite fermions. Each of these fermions behaves, in a sense, as if it has captured an even number $q$ of flux quanta. Thus, the CFs feel a reduced number of external flux quanta $N_{\phi}^*=N_{\phi}-qN$, and the effective filling factor $\nu^*$ for this system is related to the true filling factor $\nu$ by the relation~:
\begin{eqnarray}
\nu&=&\frac{\nu^*}{1+q\nu^*}.
\label{nunustar}
\end{eqnarray}
When the CFs completely occupy an integer number of pseudo-Landau levels, i.e., when $\nu^*$ is an integer $p$, they produce incompressible states at the Jain's principal sequences $\nu=p/(qp+1)$, which includes the Laughlin states at $\nu=1/(q+1)$.
The CF theory also guides construction of trial wave functions. They are of the form $\Psi^f={\cal P}_{\rm LLL}\phi^q_1 \phi_{\nu^*}$ where $\phi_{\nu^*}$ is the Slater determinant corresponding to the uncoupled CFs and $\phi_1$ is the filled Landau level (i.e. the Jastrow factor)~:
\begin{eqnarray}
\phi_1&=&\prod_{i<j}\left(z_i-z_j\right)\;,\label{jastrowfactor}
\end{eqnarray}
which encodes the flux attachment. Because the product $\phi^q_1 \phi_{\nu^*}$ has components in higher Landau levels, it is projected onto the LLL, as indicated by the projection operator ${\cal P}_{\rm LLL}$.

The CF construction trivially generalizes for bosons, with the slight change that $q$ is now taken as an odd integer. The CFs built from this process obey fermionic statistics~\cite{Leinaas77}, which means we can use the same construction as in the electronic case (i.e. filling pseudo Landau levels). So for bosons, the principal sequence occurs at~:
\begin{eqnarray}
\nu&=&\frac{p}{p+1},
\label{jainbosons}
\end{eqnarray}
and the corresponding trial wave functions are $\Psi^b={\cal P}_{\rm LLL}\phi_1 \phi_{\nu^*}$.
Trial wave functions can also be written for the spherical geometry~\cite{Jain97,Chiachen05}. For the ground state associated to the fraction $\nu=p/(p+1)$, the relation between $S$ and $N$ is given by
\begin{eqnarray}
2S&=&\frac{p+1}{p} N - p - 1.
\label{SNrelationCF}
\end{eqnarray}
In general, the trial wave functions are more complicated than Laughlin's, but many of their essential properties can be explicitly derived, such as the maximum angular momentum of the excitons, or that the ground state has $L=0$ total angular momentum (which follows from the fact that the ground state contains filled shells of composite fermions).

\section{Evidence from exact diagonalization}\label{chnumerical}

\subsection{spectrum analysis}\label{chspectrum}

\begin{figure}[!htbp]
\begin{center}
\includegraphics[width=5cm, keepaspectratio]{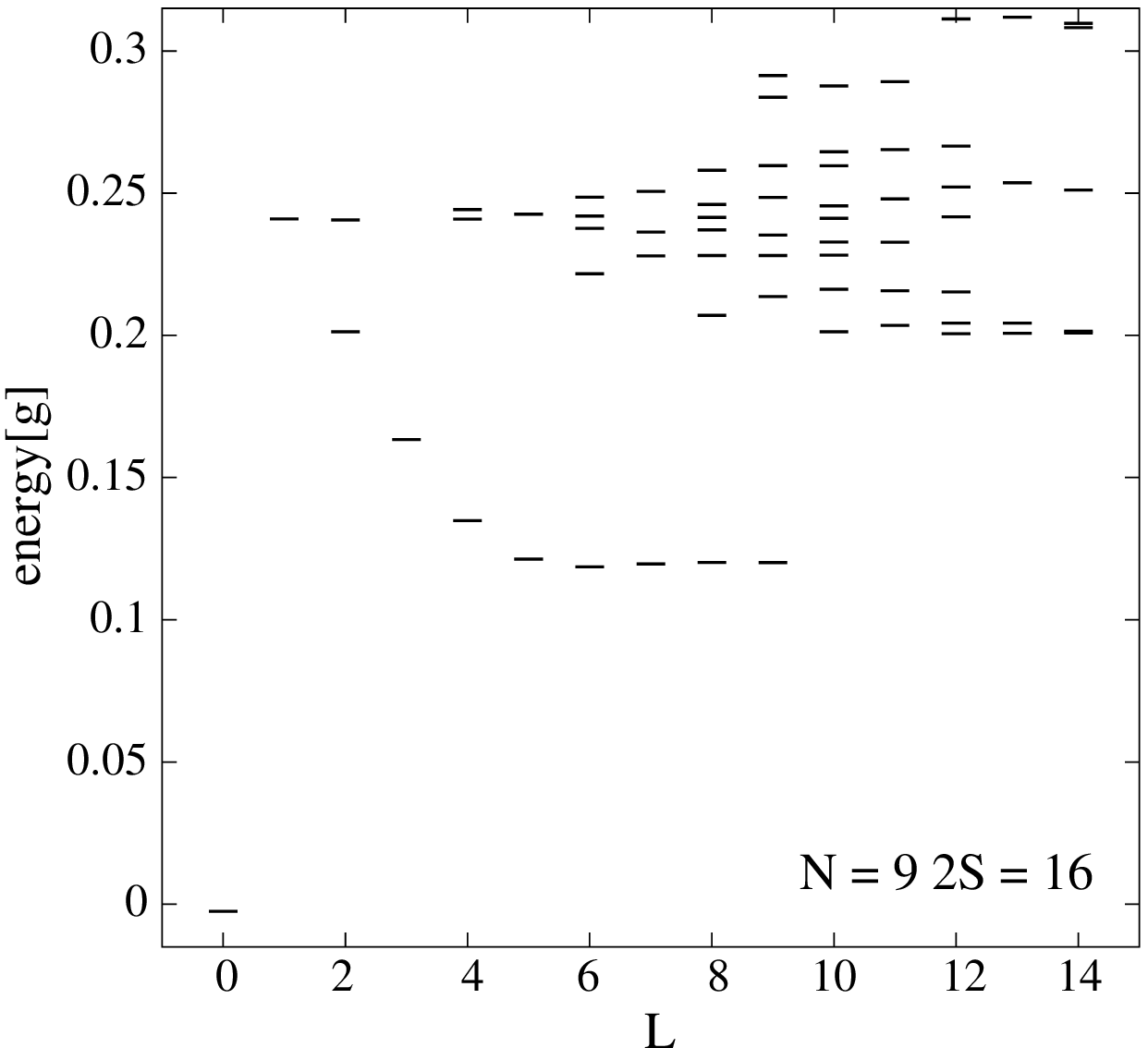}
\includegraphics[width=5cm, keepaspectratio]{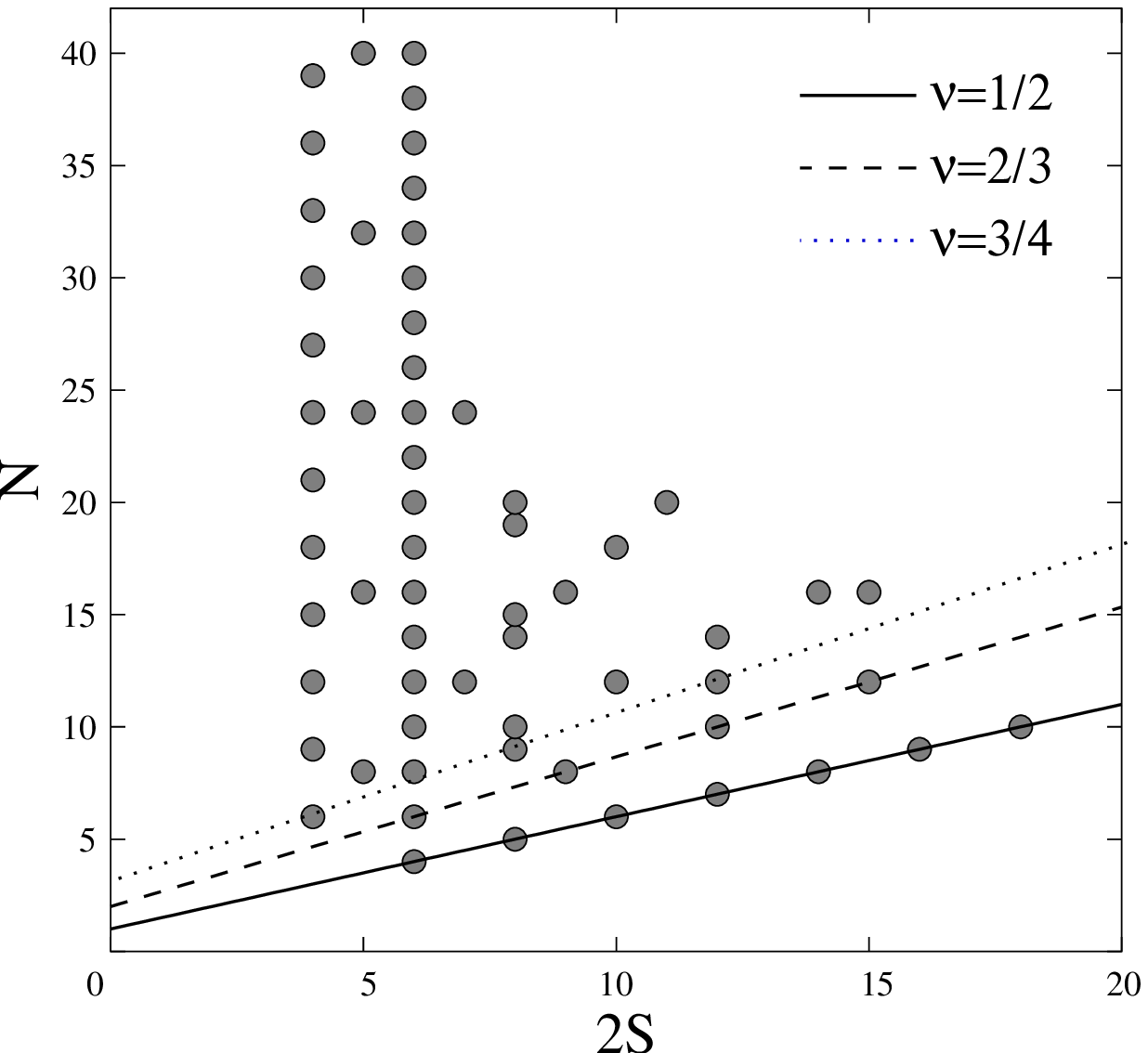}
\end{center}
\caption{{\it Left panel} : The low energy spectrum of a typical incompressible state in  the spherical geometry (here for $N=9$ at $\nu=1/2$). The ground state at $L=0$ (in this case the Laughlin state) is clearly separated from the excitations. {\it Right panel} : The candidates for incompressible states are marked on the $N-2S$ plane, where $N$ is the  number of bosons and $2S$ the number of flux quanta through the surface of the sphere. Each fraction corresponds to a line with a non-trivial shift (plotted here for fractions $\nu=1/2$, $\nu=2/3$ and $\nu=3/4$).}
\label{incompressible}
\end{figure}

Exact diagonalization has been widely used to investigate the FQHE since its early days. Even for a small number of particles, clear signatures of FQHE are exhibited in numerical data~\cite{Haldane85}. In our calculations, we use the spherical geometry and diagonalize the Hamiltonian in a fixed $L_z$ subspace.  Because we are mainly interested in the low energy physics, numerical diagonalization is performed using the L\'anczos algorithm, which allows a determination of eigenvectors and eigenvalues for large matrices.  The largest Hilbert space dimensions we can handle are of the order of $10^7$, which enables us to deal with up to $N\simeq 15-30$ particles, depending on the filling factor.   Candidates for incompressible states are identifiable by  their spectrum, because they have a clearly non-degenerate ground state (with the total orbital angular momentum $L=0$). Figure~\ref{incompressible} shows the spectrum of a typical such spectrum. Once all reachable spectra have been evaluated, we start searching for sets of $(2S,N)$ values that satisfy a linear relation of the type~:
\begin{eqnarray}
2S&=&\frac{1}{\nu} N - {\rm shift}.
\label{SNrelation}
\end{eqnarray}
If all $(2S,N)$ obeying this relation are incompressible, {\em and} the gap extrapolates to a non-zero value in the thermodynamic limit, then the corresponding fraction should exhibit FQHE in the real system. We find this to be the case for fractions $\nu=1/2$, $2/3$ and $3/4$~\cite{Regnault03}. For these three fractions, the relation between $2S$ and $N$ is the one predicted by Eq.(\ref{SNrelationCF}) (including the non-trivial shift). This is a strong hint that the CF theory correctly describes these fractions. Other spectral evidences such as the maximum angular momentum of collective mode, or the $L$ value of the quasiparticule excitations, further reinforce the CF hypothesis. But the most direct and compelling verification is provided by the overlap calculations.

\subsection{Overlap calculations}\label{choverlap}

\begin{figure}[!htbp]
\begin{center}
\includegraphics[width=5cm, keepaspectratio]{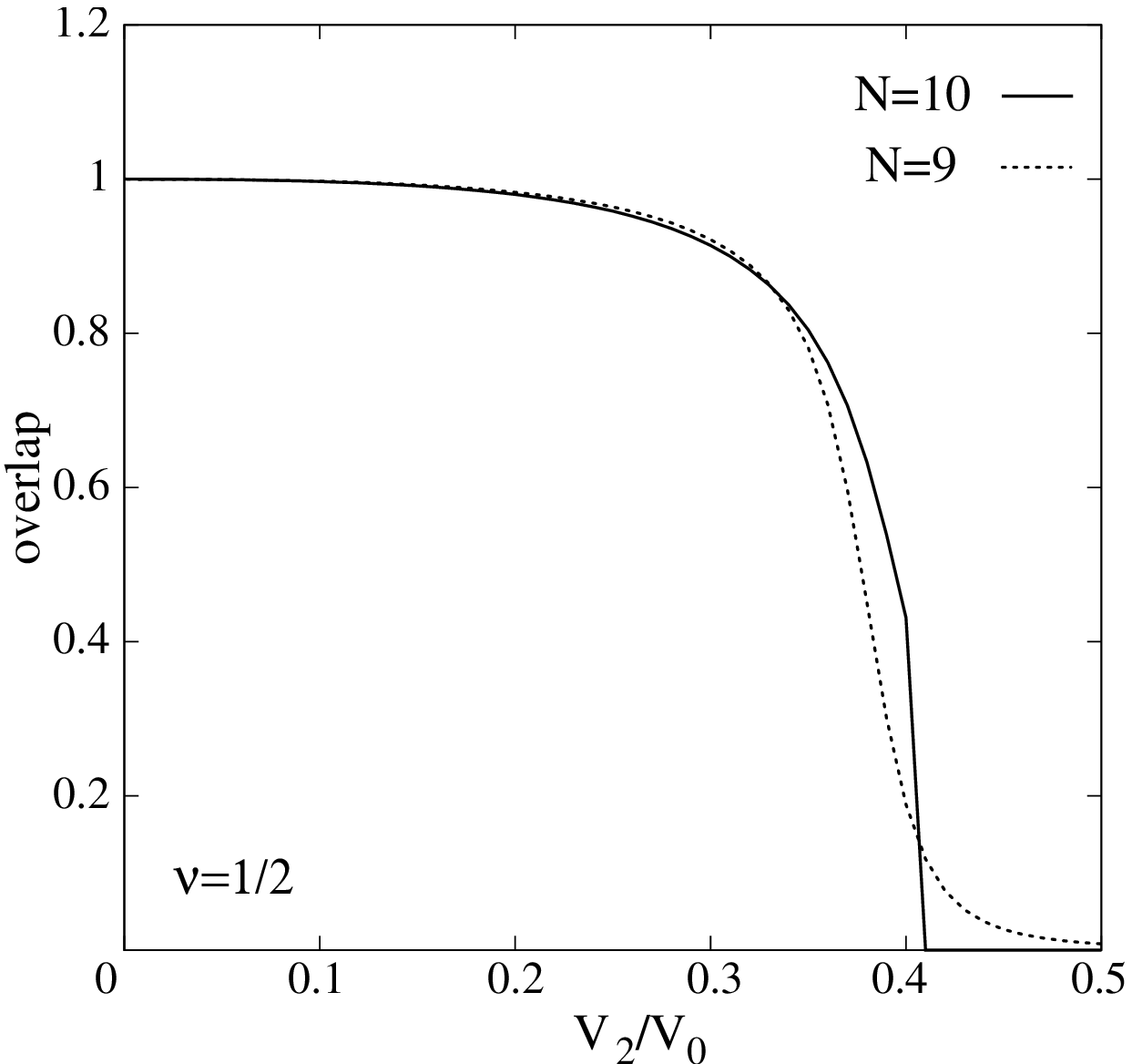}
\includegraphics[width=5cm, keepaspectratio]{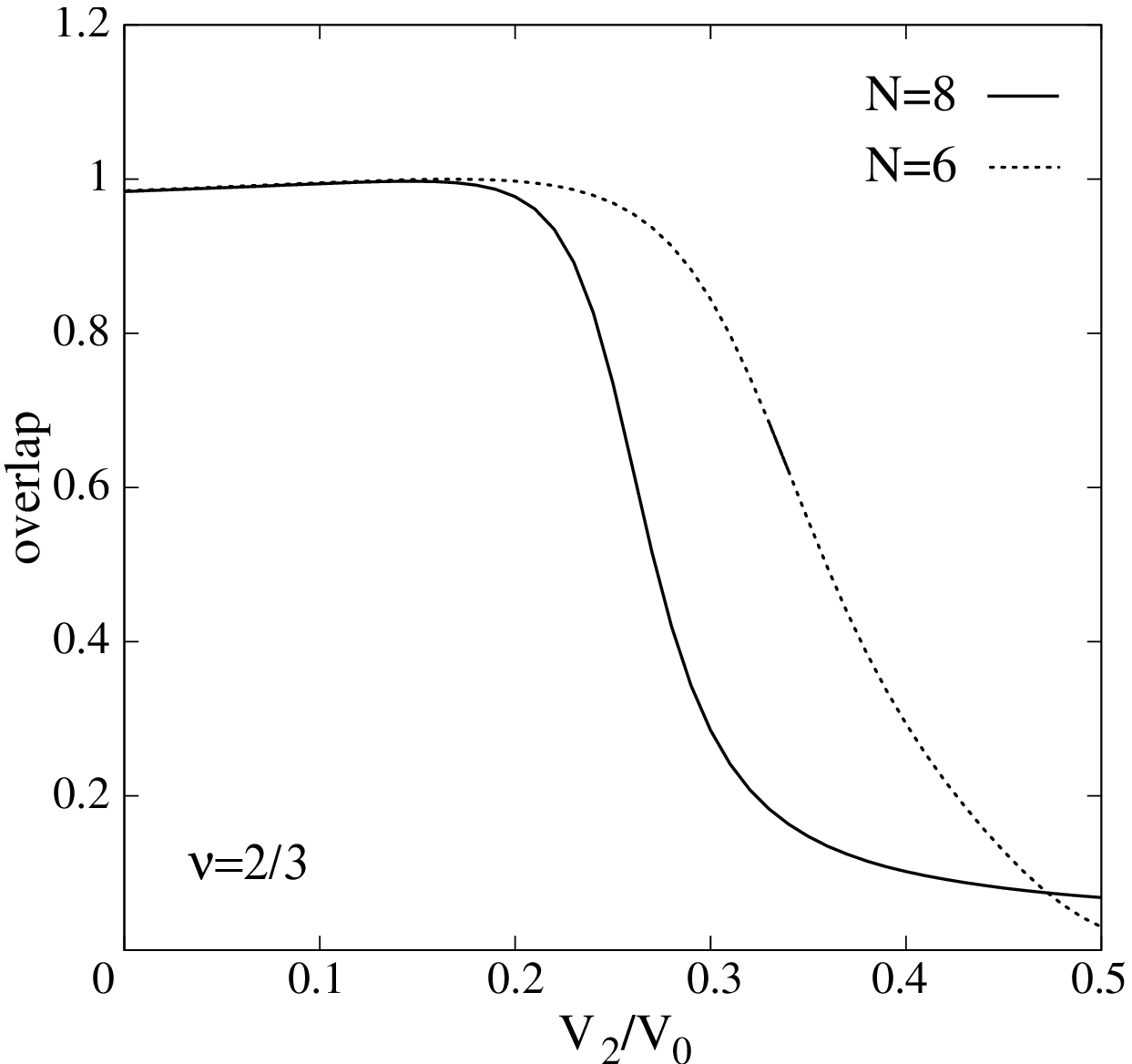}
\end{center}
\caption{Overlap of the exact ground state of an effective Hamiltonian, with an additional longer range component $V_2$ in addition to the contact interaction $V_0$, with Laughlin's and Jain's trial wave functions at $\nu=1/2$ (left panel) and $\nu=2/3$ (right panel), as a function of $V_2/V_0$.  The overlaps in the left panel are evaluated exactly, and those in the right panel by Monte Carlo method (statistical error not shown).}
\label{overlapcfv2v0}
\end{figure}

The strength of Jain's CF theory is that it predicts not only the filling factors for incompressibility but also provides explicit wave functions for ground and excited states. Numerical comparisons can be carried out between these wave functions and those obtained from exact diagonalization. The overlap between the exact state $\Psi_{{\rm ex}}$ and the trial wave function $\Psi_{{\rm trial}}$ is defined as follows~:
\begin{eqnarray}
{\cal O}&=&\frac{\left|\langle\Psi_{{\rm ex}}|\Psi_{{\rm trial}}\rangle\right|^2}{\left|\langle\Psi_{{\rm ex}}|\Psi_{{\rm ex}}\rangle\right|\left|\langle\Psi_{{\rm trial}}|\Psi_{{\rm trial}}\rangle\right|}.
\label{overlapdefinition}
\end{eqnarray}
Scalar products are calculated using Monte Carlo integration. This requires an evaluation of the exact state in the real space, involving a computation of a huge number of permanents, which is a time consuming task (see \cite{Chiachen05} for more details).
Results of overlap calculations~\cite{Chiachen05} are displayed in table \ref{jainoverlap} for fractions $\nu=1/2$ and $2/3$ and for both ground and excited states. They prove that the CF theory is in good agreement with exact diagonalizations. The CF description is also robust with respect to changes of the interaction. Figure \ref{overlapcfv2v0} shows the evolution of the ground state when we tune the interaction from pure delta function  to a longer range interaction by adding some $V_2$ pseudo-potential term. Obviously the CF approach is valid for a wide range of interactions (including the Coulomb interaction~\cite{Chiachen05}). For other fractions of the principal sequence $p/(p+1)$, the situation becomes worse as $p \longrightarrow \infty$, i.e. as $\nu$ approaches $1$. A more careful study of this limit remains to be done.

\begin{table}
\caption{\label{jainoverlap} Overlaps between exact wave functions and trial wave functions for ground and excited states for various system sizes, denoted by ${\cal O}_{{\rm gr}}$ and ${\cal O}_{{\rm ex}}$. For $\nu=1/2$, the trial wave function for the ground state is Laughlin's wave function, and the trial wave functions for the first excited states at angular momentum $L$ are Jain's wave functions. For $\nu=2/3$, both the ground state and the first excited state overlaps involve the Jain's wave functions. The statistical uncertainty in the last two digits from the Monte Carlo integration is shown in parentheses when it is larger than $10^{-5}$.}
\begin{indented}
\item[]\begin{tabular}{@{}ccccccccccc}
\br
$\nu$ & $N$ & ${\cal O}_{{\rm gr}}$ & $L$ & ${\cal O}_{{\rm ex}}$ & & $\nu$ & $N$ & ${\cal O}_{{\rm gr}}$ & $L$ & ${\cal O}_{{\rm ex}}$\\
\mr
$1/2$ & $4$ & $1.0000$ & $4$ & $0.9972$ & & $2/3$ & $4$ & $1.0000$ & $2$ & $1.0000$ \\
      & $5$ & $1.0000$ & $4$ & $0.9965$ & &       & $6$ & $0.9850$ & $4$ & $0.7544(05)$ \\
      & $6$ & $1.0000$ & $5$ & $0.9959$ & &       & $8$ & $0.9820(10)$ & $5$ & $0.8701(14)$ \\
      & $7$ & $1.0000$ & $5$ & $0.9954$ & &       & $10$ & $0.9724(89)$ & $6$ & $0.855(12)$ \\
      & $8$ & $1.0000$ & $6$ & $0.9945$ & &       &  &  &  &  \\
      & $9$ & $1.0000$ & $6$ & $0.9954(2)$ & &       &  &  &  &  \\
\br
\end{tabular}
\end{indented}
\end{table}

\section{$\nu=1$: the Pfaffian state}\label{chpfaffian}

The 2DES does not show FQHE at $\nu=1/2$~\cite{Willet93}. It is the accumulation point of the Jain's principal sequence $\nu=p/(2p+1)$, and exhibits behavior akin to a Fermi sea. This property can be understood naturally within the CF theory, as the effective magnetic field vanishes at $\nu=1/2$. Thus CFs behave like free fermions in a zeroth-order approximation.

\begin{figure}[!htbp]
\begin{center}
\includegraphics[width=5cm, keepaspectratio]{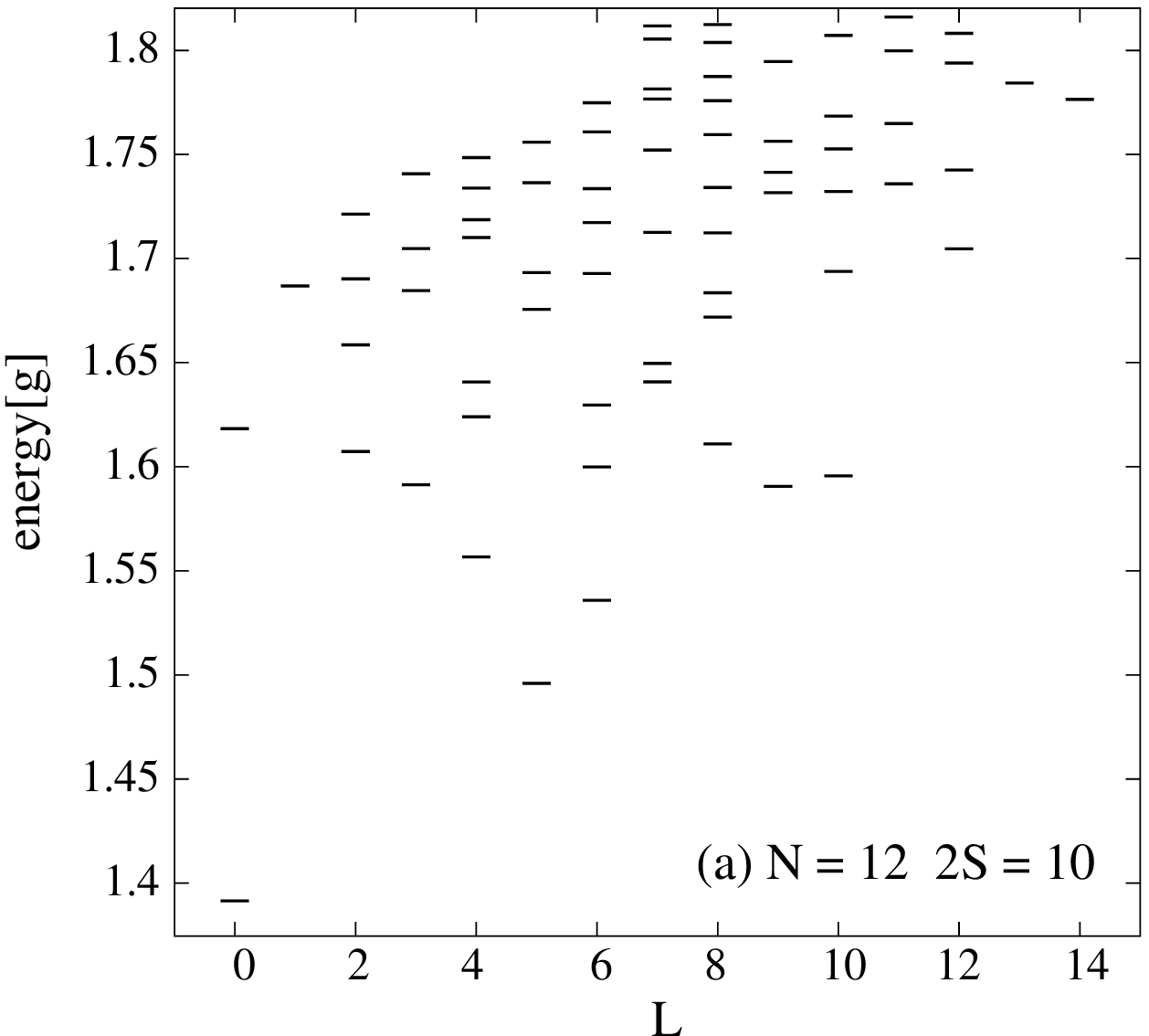}
\includegraphics[width=5cm, keepaspectratio]{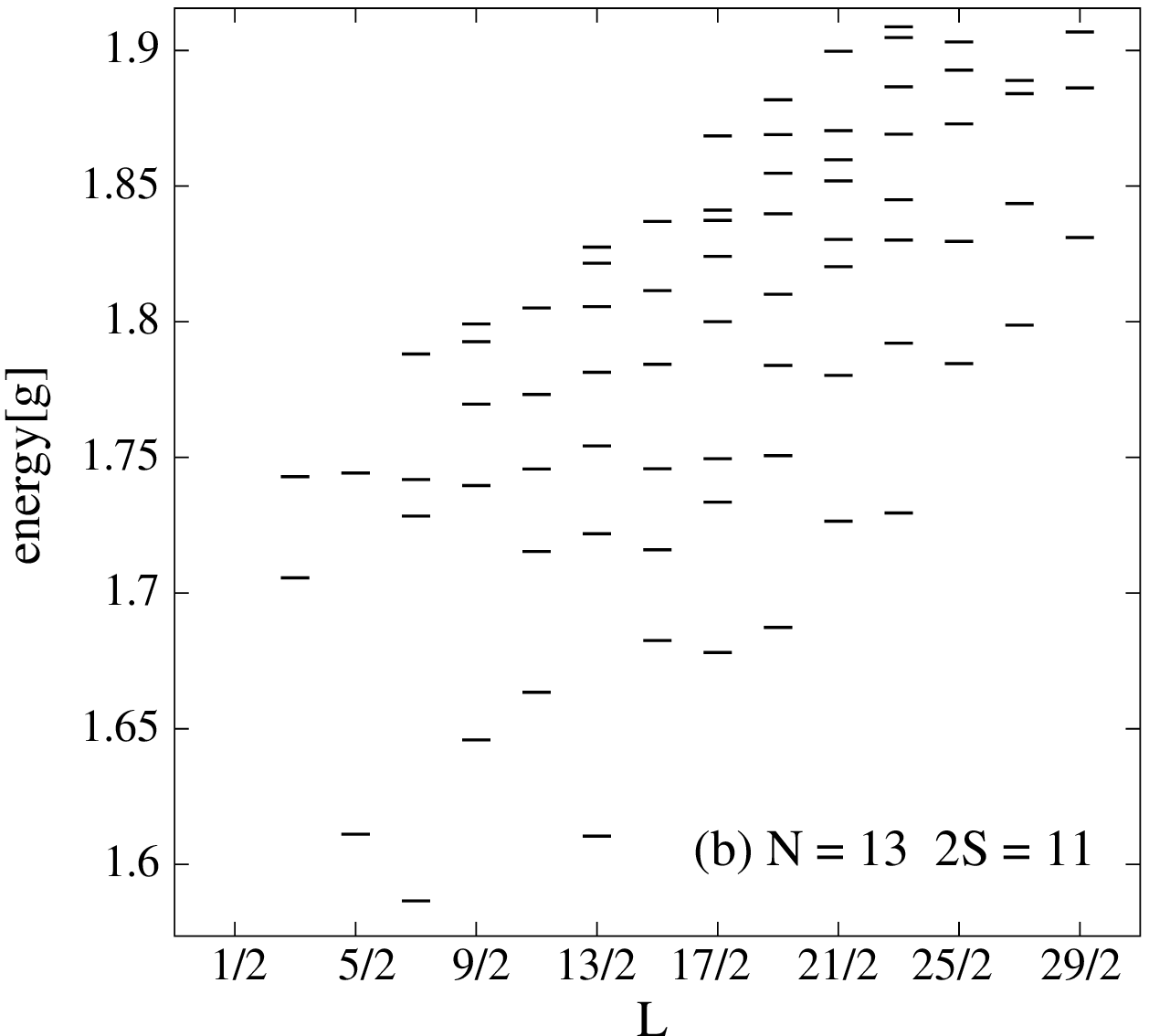}
\includegraphics[width=5cm, keepaspectratio]{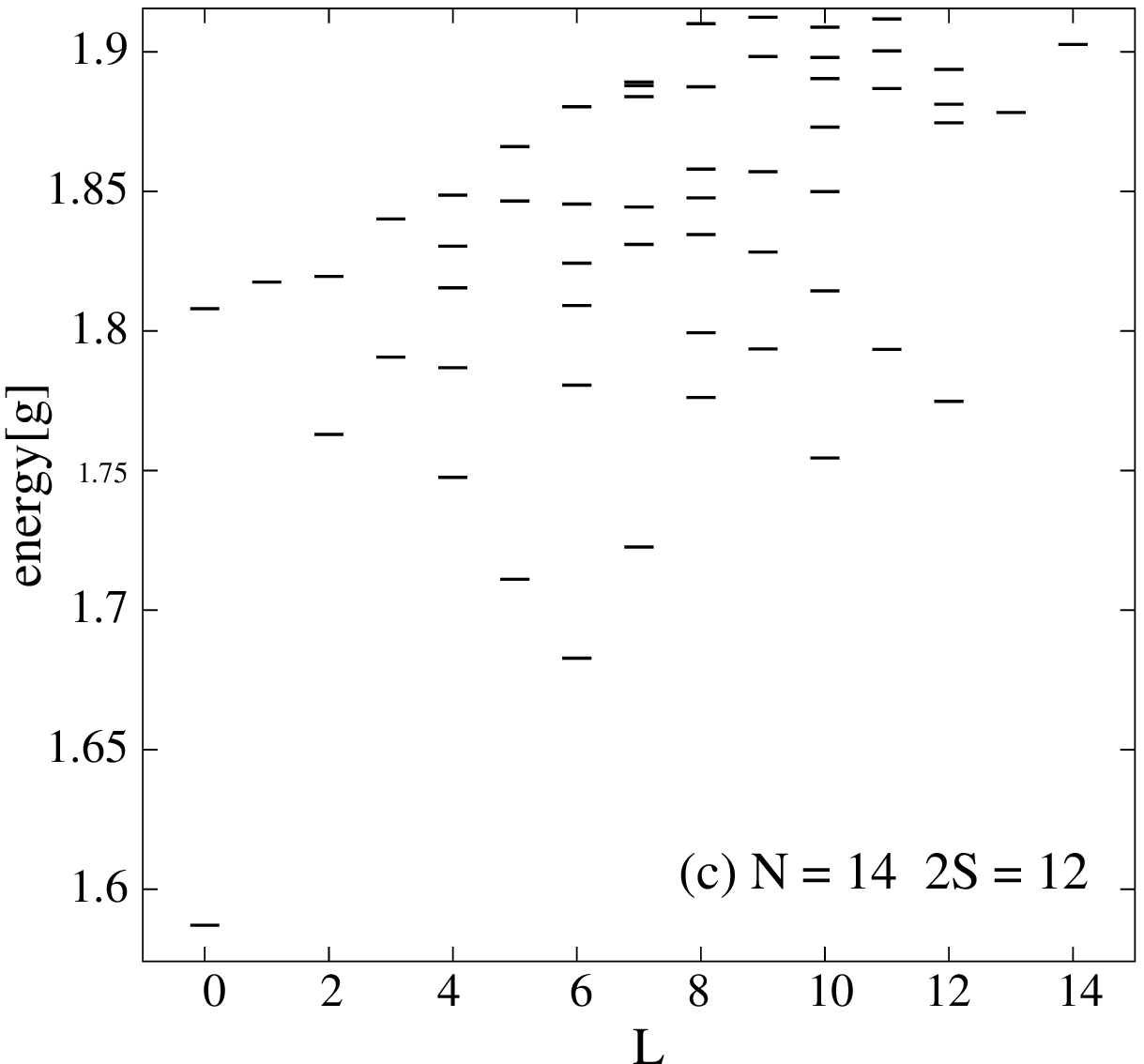}
\end{center}
\caption{Spectra at filling factor $\nu=1$ for $N=12, 2S=10$ (a), $N=13, 2S=11$ (b) and $N=14, 2S=12$ (c) in the spherical geometry. The incompressible states only appear for even numbers of particles. The relation between the number of bosons and flux quanta for these incompressible states is the one predicted by the Moore-Read wave function.}
\label{pfaffiandiag}
\end{figure}

The obvious question is then whether a Fermi sea of composite fermions occurs for bosons at the corresponding filling factor $\nu=1$.  This would be true if the model of free composite fermions is qualitatively valid here.  It has been argued the residual interaction can lead to pairing of CFs\cite{Scarola00}. This proposal has been introduced to explain the incompressible state observed at $\nu=5/2$ in 2DES. Moore and Read\cite{Moore91} have proposed an explicit trial wave function to describe the paired ground state which is analogous to the BCS wave function in real space for a given number of particles. The corresponding wave function for bosons can be written for the spherical geometry as~:
\begin{eqnarray}
\Psi_{{\rm Pf}}\left(z_1,...,z_N\right)&=&{\rm Pf}\left(\frac{1}{u_i v_j - u_j v_i}\right)\prod_{i<j}\left(u_i v_j - u_j v_i\right),\label{pffafianstate}
\end{eqnarray}
where ${\rm Pf}(A)$ is the Pfaffian of the skew-symmetric $N\times N$ matrix $A$ ($N$ even), defined as ~:
\begin{eqnarray}
{\rm Pf}\left(A\right)&=&\sum_{\sigma} \epsilon_{\sigma} A_{\sigma(1)\sigma(2)}A_{\sigma(3)\sigma(4)}...A_{\sigma(N-1)\sigma(N)},
\label{pfaffiandefinition}
\end{eqnarray}
where the sum runs over all permutations and $\epsilon_{\sigma}$ is the signature of the permutation. If such a state is present, it should appear for even numbers of particles at $2S=N-2$. Exact diagonalization clearly shows incompressible states at these values of $(2S,N)$ and also parity effect (see figure \ref{pfaffiandiag}). Thermodynamic  extrapolation of the gap leads to a non-zero value \cite{Regnault03}.

\begin{figure}[!htbp]
\begin{center}
\includegraphics[width=5cm, keepaspectratio]{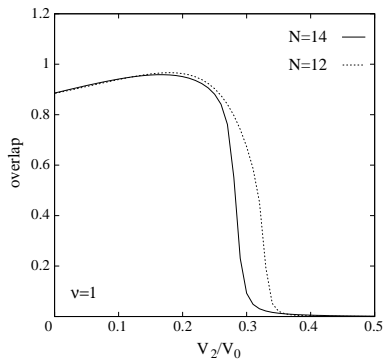}
\end{center}
\caption{Overlap of the Pfaffian state with the exact ground state of the effective Hamiltonian with an additional longer range component $V_2$. The overlaps are calculated for $N=14$ and $N=12$ bosons. The description in terms of the Pfaffian state is robust with respect to long range interaction up to a critical value of $V_2/V_0$.}
\label{overlappfaffianv2v0}
\end{figure}

Direct comparison between the exact ground state and the Moore-Read (or Pfaffian) wave function can also be performed~\cite{Chiachen05}. The latter can be generated by exact diagonalization using the fact that Eq.(\ref{pffafianstate}) is the exact zero-energy ground state of the three-body hardcore Hamiltonian~\cite{Greiter91,Read99}~:
\begin{eqnarray}
{\cal{H}}_{{\rm Pf}}&=&\sum_{i<j<k}\delta^{(2)}\left(u_i v_j - u_j v_i\right)\delta^{(2)}\left(u_j v_k - u_k v_j\right).
\label{threebodyhamiltonian}
\end{eqnarray}
The overlaps are thus obtained by a simple scalar product of two ground 
state vectors. Overlap results in table \ref{pfaffianoverlap} show that the description based on the Pfaffian wave function is valid for the delta function interaction; figure \ref{overlappfaffianv2v0} demonstrates its robustness for longer range interactions as well.

\begin{table}
\caption{\label{pfaffianoverlap} Overlap between the exact ground state at $2S=N-2$ and the Pfaffian state for different system sizes.}
\begin{indented}
\item[]\begin{tabular}{@{}lccccccc}
\br
$N$&$4$&$6$&$8$&$10$&$12$&$14$&$16$\\
\mr
${\cal O}$&$1.0$&$0.9728$&$0.9669$&$0.9592$&$0.8844$&$0.8858$&$0.8833$\\
\br
\end{tabular}
\end{indented}
\end{table}

\section{Conclusion}

Both analytical and numerical results suggest that ultracold boson atoms in rotating trap should display a very rich physics in the low filling factor regime. The analogy with the FQHE in 2DES suggests the use of the CF approach for the bosonic system, and indeed, numerical tests prove its validity. The robustness of the CF description for fractions $\nu=1/2$ and $\nu=2/3$ is clearly demonstrated. Nevertheless, there are some major differences from the 2DES systems. The residual interactions between the composite fermions tend to complicate the description of fractions $p/(p+1)$ with increasing $p$, leading to a paired state at $\nu=1$ well described by a BCS-like Pfaffian wave function. 
Also, in this paper, we do not address the $\nu>1$ case.  CF theory allows construction of the sequence $\nu=p/(p-1)$ by reversing the attached flux quanta. But the overlap calculations tend to indicate that this description is not valid (at least for $\nu=2$ and $\nu=3/2$). Other states have been predicted \cite{Cooper01} to appear at filling factors $\nu=k/2$ ($k$ being an integer value), correspond to Read-Rezayi states. They are generalizations of the Pfaffian state and also exhibits excitations obeying non-Abelian statistics. But it is not yet clear whether this description is valid in the thermodynamic limit for the delta function interaction~\cite{Regnault03}. Numerical evidence~\cite{Rezayi05} indicates that longer range interaction (like dipole-dipole interaction) help to stabilize such exotic states.

\ack

NR wishes to thank the BEC 2005 organizers and committee for their invitation. Partial support of this research by the National Science Foundation under grant no. DMR-0240458 is acknowledged. We are grateful for a computer time allocation of IDRIS-CNRS and to the High Performance Computing (HPC) Group for assistance and computing time with the LION-XL cluster.


\section*{References}
\begin{thebibliography}{99}

\bibitem{Girardeau60} M. Girardeau, J. Math. Phys. {\bf 1}, 516 (1960).

\bibitem{Lieb63} E. Lieb and W. Liniger, Phys. Rev. {\bf 130}, 1605 (1963); C.N. Yang and C.P. Yang, J. Math. Phys. {\bf 10}, 1115 (1969).

\bibitem{Kinoshita04} T. Kinoshita, T. Wenger, and D.S. Weiss, Science {\bf 305}, 1125 (2004).

\bibitem{Paredes04} B. Paredes {\em et al.} Nature {\bf 429}, 277 (2004).

\bibitem{Madison00} K. W. Madison , F. Chevy, W. Wohlleben, and J. Dalibard, Phys.Rev. Lett. {\bf 84}, 806 (2000); F. Chevy, K. Madison, and J. Dalibard, Phys. Rev. Lett. {\bf 85}, 2223 (2000).

\bibitem{Aboshaeer01} J. R. Abo-Shaeer , C. Raman, J. M. Vogels , and W. Ketterle, Science {\bf 292}, 476 (2001).

\bibitem{Wilkin98} N.K. Wilkin, J.M.F. Gunn and Smith, Phys. Rev. Lett. {\bf 80}, 2265 (1998).

\bibitem{Wilkin00} N.K. Wilkin and J.M.F. Gunn, Phys. Rev. Lett. {\bf 84}, 6 (2000).

\bibitem{Cooper01} N.R. Cooper, N.K. Wilkin, and J.M.F. Gunn, Phys. Rev. Lett. {\bf 87}, 120405 (2001).

\bibitem{Regnault03} N. Regnault and Th. Jolicoeur, Phys. Rev. Lett. {\bf 91}, 030402 (2003); Phys. Rev. B {\bf 69}, 235309 (2004).

\bibitem{Regnault04} N. Regnault and Th. Jolicoeur, Phys. Rev. B {\bf 70}, 241307 (2004).

\bibitem{Baranov05} M. A. Baranov, K. Osterloh, and M. Lewenstein, Phys. Rev. Lett. {\bf 94}, 070404 (2005).

\bibitem{Laughlin83} R.B. Laughlin, Phys. Rev. Lett. {\bf 50}, 1395 (1983).

\bibitem{Jain89}  J.K. Jain, Phys. Rev. Lett. {\bf 63}, 199 (1989); Physics Today {\bf 53}(4), 39 (2000); Physica E {\bf 20}, 79 (2003).

\bibitem{Wilczek90} F. Wilczek, {\it ``Fractional Statistics and Anyon Superconductivity''}, World Scientific Singapore (1990).

\bibitem{Xie91} X.C. Xie, S. He, and S. Das Sarma, Phys. Rev. Lett. {\bf 66}, 389 (1991).

\bibitem{Quinn01} J.J. Quinn, A. Wojs, J.J. Quinn and A. Benjamin, Physica E {\bf 9}, 701 (2001).

\bibitem{Viefers00} S. Viefers, T.H. Hansson, S.M. Reimann, Phys. Rev. A {\bf 62}, 053604 (2000).

\bibitem{Cooper99} N.R. Cooper and N.K. Wilkin, Phys. Rev. B {\bf 60}, R16279 (1999).

\bibitem{Wu76} T. T. Wu and C. N. Yang, Nucl. Phys. B{\bf 107}, 365 (1976); Phys. Rev. D{\bf 16}, 1018 (1977).

\bibitem{Haldane83} F. D. M. Haldane, Phys. Rev. Lett. {\bf 51}, 605 (1983).

\bibitem{Fano86} G. Fano, F. Ortolani, and E. Colombo, Phys. Rev. B {\bf 34}, 2670 (1986).

\bibitem{Leinaas77} J.M. Leinaas and J. Myrheim, Nuovo Cimento B {\bf 37}, 1 (1977); F. Wilczek, Phys. Rev. Lett. {\bf 48}, 1144 (1982).

\bibitem{Jain97} J.K. Jain and R.K. Kamilla, Int. J. Mod. Phys. B {\bf 11}, 2621 (1997); Phys. Rev. B {\bf 55}, R4895 (1997).

\bibitem{Chiachen05} C. Chang, N. Regnault, Th. Jolicoeur and J. K. Jain, Phys. Rev. A {\bf 72}, 013611 (2005).

\bibitem{Haldane85} F. D. M. Haldane and E. H. Rezayi, Phys. Rev. Lett. {\bf 54}, 237 (1985).

\bibitem{Willet93} R.L. Willet, {\em et al}. Phys. Rev. Lett. {\bf 71}, 3846 (1993); V.J. Goldman, {\em et al}. Phys. Rev. Lett. {\bf 72}, 2065 (1994); W. Kang, {\em et al}. Phys. Rev. Lett. {\bf 71}, 3850 (1993). 

\bibitem{Scarola00} V.W. Scarola, K. Park, and J.K. Jain, Nature {\bf 406}, 863 (2000).

\bibitem{Moore91} G. Moore and N. Read, Nucl. Phys. B {\bf 360}, 362 (1991).

\bibitem{Greiter91} M. Greiter, X.-G. Wen, and F. Wilczek, Phys. Rev. Lett. {\bf 66}, 3205 (1991).

\bibitem{Read99} N. Read and E. Rezayi, Phys. Rev. B {\bf 59}, 8084 (1999).

\bibitem{Rezayi05} E. Rezayi, N. Read and N.R. Cooper, Phys. Rev. Lett. {\bf 95}, 160404 (2005).

\end{thebibliography}
\end{document}